\documentclass[mypaper,7pt,twoside]{CoAst}
\usepackage{epsf,graphicx,fancyhdr}
\renewcommand{\chaptermark}[1]{\markboth{#1}{}}

\renewcommand{\headrulewidth}{0pt}
\newcommand{\apj}{ApJ}
\newcommand{\aap}{A\&A}  
\newcommand{\mnras}{MNRAS}

\newcommand{\kopf}{\small\itshape Comm. in Asteroseismology\\ Vol. 148, 2007}
\newcommand{\Authors}[1]{\begin{center}\normalsize\bf\sf #1 \end{center}}

\renewcommand{\author}[1]{\begin{center}\normalsize\bf\sf #1 \end{center}}
\newcommand{\Address}[1]{\begin{center}\small\sf #1 \end{center}}

\renewenvironment{abstract}{\section*{Abstract}\normalsize\sf}{}
\newcommand{\References}[1]{\begin{flushleft}{\large References\\}\vspace*{2mm}\small #1 \end{flushleft}}

\newcommand{\chapterDSSN}[2]{\chapter[\sf\normalsize #1\\ \footnotesize \hspace*{5mm}by #2 \sf\normalsize][]{#1\\}\rhead[\fancyplain{}{\sf\footnotesize \center{#1}}]{\fancyplain{}{\sffamily\thepage}}\lhead[\fancyplain{\kopf}{\sffamily\thepage}]{\fancyplain{\kopf}{\sf\footnotesize \center{#2}}}}

\newcommand{\figureDSSN}[5]{\begin{figure}[#4]
\centering
\includegraphics*[#5]{#1}
\caption{#2}
\label{#3}
\end{figure}}

\renewcommand{\chaptername}{}
\newcommand{\acknowledgments}[1]{\vspace*{5mm}\noindent\begin{bf}Acknowledgments. \end{bf} #1}

\begin{document}
\sf

\chapterDSSN{The Future of Computational Asteroseismology}{T.~S.~Metcalfe}

\Authors{Travis S.~Metcalfe} 
\Address{High Altitude Observatory \& Scientific Computing Division,\\
National Center for Atmospheric Research, P.O.~Box 3000, Boulder CO 80307 USA}

\noindent
\begin{abstract}
The history of stellar seismology suggests that observation and theory 
often take turns advancing our understanding. The recent tripling of the 
sample of pulsating white dwarfs generated by the Sloan Digital Sky Survey 
represents a giant leap on the observational side. The time is ripe for a 
comparable advance on the theoretical side. There are basically two ways 
we can improve our theoretical understanding of pulsating stars: we can 
improve the fundamental ingredients of the models, or we can explore the 
existing models in greater computational detail. For pulsating white 
dwarfs, much progress has recently been made on both fronts: models now 
exist that connect the interior structure to its complete evolutionary 
history, while a method of using parallel computers for global exploration 
of relatively simple models has also been developed. Future advances in 
theoretical white dwarf asteroseismology will emerge by combining these 
two approaches, yielding unprecedented insight into the physics of 
diffusion, nuclear burning, and mixing.
\end{abstract}


\section*{1. Context} 

In just the past few years, the Sloan Digital Sky Survey has {\it 
tripled\,} the sample of pulsating white dwarf stars (Mukadam et 
al.~2004, Mullally et al.~2005, Kepler et al.~2005, Castanheira et 
al.~2006). In the next few years we can expect similar increases to emerge 
for other types of pulsating stars, from space missions such as CoRoT and 
Kepler. The observations are quickly becoming too numerous for us to do 
traditional model-fitting by hand. We need to automate the procedure so we 
can spend more of our time thinking about the results.

The great thing about computers is that they can work 24 hours a day, and 
they are so inexpensive that you can have many of them working for you at 
once (Metcalfe \& Nather 2000). There are many exciting areas of pulsating 
star research where analytical work can contribute much to our 
understanding, but I won't discuss them here. This will be a 
computer-centric view of the future. After a brief overview of white dwarf 
asteroseismology (\S2), I will outline two broad approaches that are 
currently being used to obtain physical insight from computational work 
(\S3 and \S4). In the future, as our computing potential increases, we 
will eventually be able to combine these two approaches (\S5).


\section*{2. White Dwarf Asteroseismology}

First let me remind you that there are several major spectroscopic classes 
of white dwarfs. With few exceptions, we expect all of them to have cores 
composed of a mixture of carbon and oxygen---the ashes of helium burning 
during the red giant phase (Metcalfe 2003). About 80\% of white dwarfs are 
classified as type DA, with a mantle of helium above the core and beneath 
a thin surface layer of pure hydrogen. Most of the remaining 20\% show no 
traces of hydrogen---exposing either a pure helium surface (type DB), or a 
mixture of helium, carbon, and oxygen (type DO). Each of these major 
spectral types produces its own class of variables in the H-R diagram, 
spanning the temperature ranges for partial ionization of carbon and 
oxygen (DOV), helium (DBV), and hydrogen (DAV). The instability strips for 
these three classes are roughly equally spaced in $\mathsf{log~T_{eff}}$
(see Elsworth \& Thompson 2004, their Fig.~1).

Like other types of pulsating stars, the spherical symmetry allows us to 
decompose the light variation into spherical harmonics, and because we do 
not have spatial resolution across the stellar surface only those modes 
with low spherical degree ($\mathsf{\ell<3}$) can be detected. In contrast 
to other types of pulsating stars, the field of white dwarf 
asteroseismology developed early along with helioseismology. This can be 
attributed to the fact that the pulsations in these stars are {\it not 
subtle\,}. The total light variations are typically $\mathsf{\sim\!10}$\% 
on convenient timescales of $\mathsf{\sim\!10}$ minutes. The simultaneous 
presence of many closely-spaced frequencies leads to easily visible 
beating in their light curves.

From a theoretical perspective, pulsation frequencies are basically 
determined by the sound speed (or Lamb frequency) and the buoyancy (or 
Brunt-V\"ais\"al\"a) frequency from the center of a star to its surface. 
Pressure modes (p-modes) are excited at relatively high frequencies 
(larger than both of these natural frequencies) while gravity modes 
(g-modes) are excited at lower frequencies, smaller than both natural 
frequencies (Unno et al.~1989). The pulsations in white dwarf stars are 
excited in the range of frequencies characteristic of g-modes, and the 
models suggest that the periods are determined primarily by the buoyancy 
frequency. The strong gravity in white dwarf stars quickly stratifies the 
surface layers, and the resulting composition gradients cause 
perturbations to the buoyancy frequency that lead to deviations from the 
uniform period spacing predicted by asymptotic theory. We can use these 
deviations to infer the interior structure through forward modeling.


\section*{3. Beyond Local Fitting}

One way that we can obtain physical insight from computational work is to 
use relatively simple models, but to explore them more globally than we 
have in the past. What does this mean in practice? Most of you are 
probably familiar with {\it Moore's Law\,}, which is really an empirical 
observation that ``computing power per unit cost doubles approximately 
every 18 months.'' This has been true for more than a century, spanning 
many different computing technologies. A somewhat less well known law 
(because I made it up) is the {\it More is Better Law\,}, which states: 
``If you spend more time writing the paper than running the models, you 
didn't run enough models.''

Of course, ``writing the paper'' is really just a euphemism for all 
aspects of a research project that do not involve either computing or 
interpreting the results. The point is that if you choose to take a 
computational approach to a problem, the actual computations should occupy 
a large fraction of the total time needed to complete the project. Thus, 
what we can accomplish is in some sense driven by {\it Moore's 
Law\,}---but we can circumvent this limitation with parallel computing, 
and then concentrate on what {\it More\,} we can do.

\subsection*{3.1 More Stars}

Although our models are often physically simplified, we can instill 
greater confidence in the computational results by fitting them to more 
stars. If this leads to a qualitatively consistent picture of what we 
theoretically expect to find, it could just be a coincidence---but as more 
and more stars support the same picture, it is easier to believe that even 
relatively simple models can provide important physical insights. Let me 
give you an example from my own work on white dwarf stars.

The DB white dwarfs are thought to evolve from hydrogen-deficient 
post-asymptotic giant branch (post-AGB) stars, which are the result of a 
very late thermal pulse leading to the so-called born-again AGB scenario 
(Iben et al.~1983). The hot DO white dwarfs that emerge initially have 
envelopes containing a uniform mixture of helium, carbon, and oxygen. As 
the DO star cools over time, the helium floats to the surface---gradually 
growing thicker and transforming the star into a DB. This process 
continues within the DB instability strip, so we can test the theory by 
measuring the thickness of the pure helium surface layer in several DBV 
stars with different temperatures.

\figureDSSN{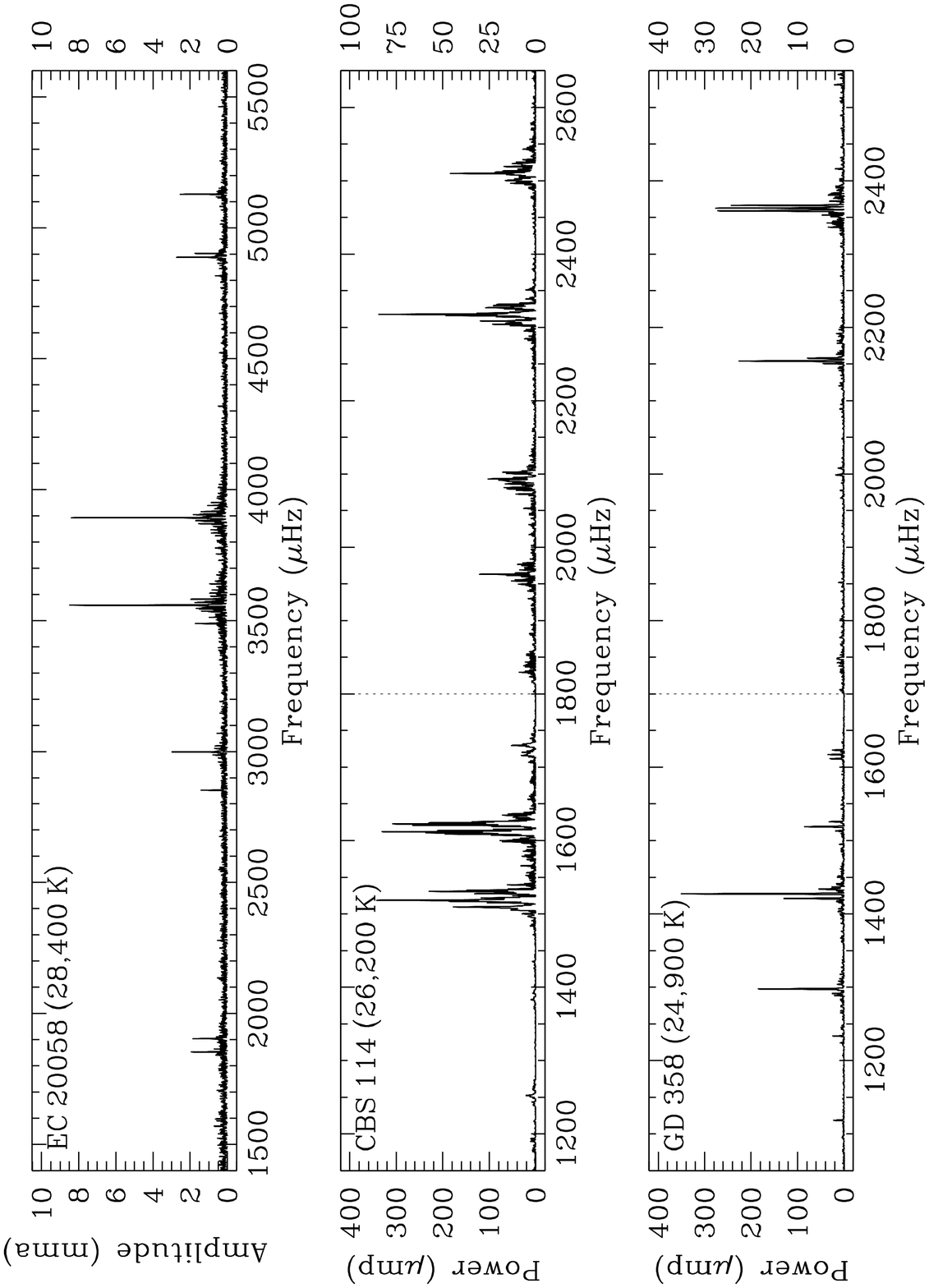}{Fourier spectra from multi-site      %
observations of three DBV white dwarfs with different temperatures  %
[data from Sullivan et al.~(in prep.), Metcalfe et al.~(2005), and  %
Winget et al.~(1994); Temperatures from Beauchamp et al.~(1999)].   %
}{fig1}{t}{clip,angle=270,width=110mm}                              %

Asteroseismic observations are now available for three DBV stars which 
exhibit many independent pulsation modes (see Figure~\ref{fig1}). Adopting 
the spectroscopic temperatures of Beauchamp et al.~(1999), the hottest of 
the stars is EC~20058 (Sullivan et al.~in prep.) at 28,400~K. Slightly 
cooler is CBS~114 (Metcalfe et al.~2005) at 26,200~K, and cooler still is 
GD~358 (Winget et al.~1994) at 24,900~K. The specific prediction of 
diffusion theory is that we should find progressively {\it thicker\,} 
surface helium layers for {\it cooler\,} DB white dwarfs.

\subsection*{3.2 More Models}

To test this prediction, we have used a parallel genetic algorithm 
(Metcalfe \& Charbonneau 2003) to globally minimize the root-mean-square 
($\mathsf{r.m.s.}$) difference between the observed and calculated 
pulsation periods in each of these three stars. We use relatively simple 
models with pure carbon cores since we are primarily interested in the 
envelope structure. The genetic algorithm searches a broad range for each 
of the four adjustable parameters, probing white dwarf masses 
($\mathsf{M_*}$) between 0.45 and 0.95~$\mathsf{M_\odot}$, effective 
temperatures ($\mathsf{T_{eff}}$) from 20,000 to 30,000~K, total envelope 
masses ($\mathsf{M_{env}}$) between $\mathsf{10^{-2}}$ and 
$\mathsf{10^{-4}~M_*}$, and surface helium layer masses 
($\mathsf{M_{He}}$) from $\mathsf{10^{-5}}$ to $\mathsf{10^{-7}~M_*}$. For 
each model-fit, the genetic algorithm calculates more than 500,000 
models---initially distributed over the full range of parameter values, 
but ultimately concentrated near the region of the global solution.

\begin{table}
\caption{Optimal model parameters for three DBV stars.\label{tab1}}
\begin{center}
\begin{tabular}{lrrrl}
\hline
\sf Parameter                      &\sf EC~20058 &\sf CBS~114 &\sf GD~358  &\sf Uncertainty     \\
\hline
$\mathsf{T_{eff}~(K)}\dotfill$     &\sf 28\,100  &\sf 25\,800 &\sf 23\,100 &$\mathsf{\pm100}$   \\
$\mathsf{M_*\ (M_{\odot})}\dotfill$&\sf 0.550    &\sf 0.630   &\sf 0.630   &$\mathsf{\pm0.005}$ \\
$\mathsf{log(M_{env}/M_*)}\ldots$  &\sf $-$3.56  &\sf $-$2.42 &\sf $-$2.92 &$\mathsf{\pm0.02}$  \\
$\mathsf{log(M_{He}/M_*)}\dotfill$ &\sf $-$6.42  &\sf $-$5.96 &\sf $-$5.90 &$\mathsf{\pm0.02}$  \\
$\mathsf{r.m.s.~(sec)}\dotfill$    &\sf 1.89     &\sf 2.33    &\sf 2.26    &$\mathsf{\cdots}$   \\
\hline
\end{tabular}
\end{center}
\end{table}

The results of the three model-fits (see Table~\ref{tab1}) are in 
qualitative agreement with the predictions of diffusion theory, with the 
inferred surface helium layer masses growing thicker for the progressively 
cooler stars. The inferred masses and temperatures for the three stars are 
in reasonable agreement with the spectroscopically determined values, and 
the total envelope masses are all within the range expected from stellar 
evolution theory (Dantona \& Mazzitelli 1979). The overall quality of each 
fit is quite good, especially considering that we have ignored any 
structure in the core.

\subsection*{3.3 More Parameters}

Of course, we {\it know\,} that real white dwarfs do not have pure carbon 
cores. The actual interior chemical profiles consist of a uniform mixture 
of carbon and oxygen out to some fractional mass that depends on the size 
of the convective core in the red giant progenitor. Outside of this 
uniform region, the oxygen mass fraction decreases to zero in a manner 
that is determined by the conditions during helium shell burning. The most 
important feature of this chemical profile, from an asteroseismic 
standpoint, is the location of the initial break from a uniform mixture of 
carbon and oxygen. But the detailed shape of the oxygen mass fraction as 
it falls to zero also matters.

We can investigate the relative importance of these effects by adding more 
parameters to our model. In the past, inferences of core structure in 
white dwarfs were made using a simple parameterization of the chemical 
profile that fixed the oxygen mass fraction to its central value 
($\mathsf{X_{O}}$) out to some fractional mass ($\mathsf{q}$) where it 
then decreased linearly to zero at the 0.95 fractional mass point 
(Metcalfe et al.~2001). Based on the calculations of Salaris et 
al.~(1997), we have recently incorporated new chemical profiles into the 
models to specify the detailed shape of the oxygen mass fraction. We use 
the same two parameters, and simply scale the shape within each model.

The initial application of these 6-parameter models to CBS~114 yields a 
globally optimal model that agrees with both the predictions of diffusion 
theory {\it and\,} the expected nuclear burning history of the progenitor 
(Metcalfe 2005). This example clearly demonstrates the potential of using 
simple models when combined with a more global fitting strategy.


\section*{4. Beyond the Spherical Cow}

At some point, even a global exploration of simple models will run into 
limitations. A broad search is certainly useful for identifying the region 
of the global solution, but the final results may suffer from small 
systematic errors in the optimal parameter values. How else might we use 
our continually expanding computational potential? Another approach is to 
build the best possible models, but limit the exploration. The analysis is 
local, but for a variety of reasons the final results may be {\it More\,} 
reliable.

\subsection*{4.1 More Physics}

The most obvious thing we can do to improve the models is simply to use 
the most accurate physical ingredients that are currently available. A 
recent example of this approach for white dwarf models can be found in 
C{\'o}rsico et al.~(2004). This study was designed to probe the 
asteroseismic differences between partially crystallized models of DAV 
stars with different core compositions. This is motivated by the fact that 
only relatively massive DA stars are expected to crystallize while still 
within the instability strip, and the transition from carbon and oxygen 
dominated cores to those containing primarily oxygen and neon takes place 
in this same mass range (Iben et al.~1997).

The authors include a time-dependent treatment of diffusion to describe 
the chemical profiles of the hydrogen and helium layers---making the 
calculations much more computationally demanding. They adopt initial 
profiles for the distribution of oxygen and neon in the cores of one set 
of models from detailed evolutionary calculations that follow the repeated 
carbon-burning shell flashes in the progenitor. For the other set of 
models with carbon and oxygen cores, they include a self-consistent 
treatment of phase separation during the crystallization process. Although 
they make no attempt to fit these models to the available observations of 
BPM~37093 (Kanaan et al.~2005), they do find significant differences in 
the pulsation properties of the two sets of models which will ultimately 
make observational tests possible.

\subsection*{4.2 More History}

We can also improve the white dwarf models by connecting them directly to 
the prior stages of their evolution. An impressive example of this 
approach was recently published by Althaus et al.~(2005), who examine the 
possible evolutionary connection between DO, DB, and DQ stars. As outlined 
briefly in section 3.1, the surface chemical composition of DO stars 
suggests that as they cool, diffusion will gradually transform them into 
DB white dwarfs. Since the surface convection zone grows deeper as the DB 
star cools further, it may eventually reach the underlying carbon-rich 
layer and dredge up enough carbon to transform the star again, this time 
into a DQ. Although there is no known class of DQ pulsators, asteroseismic 
tests of these evolutionary calculations are possible within both the DOV 
and DBV instability strips.

The study follows the complete evolution of a $\mathsf{2.7~M_\odot}$ star 
from the zero age main sequence, through mass loss on the AGB, and into 
the white dwarf regime. The authors employ a coupled treatment of nuclear 
burning and mixing, including five chemical time steps for each evolution 
step, which is especially important to follow the fast evolutionary phases 
like the born-again episode. They also adopt the double-diffusive 
mixing-length theory of Grossman \& Taam (1996) to allow non-instantaneous 
mixing for a fluid with composition gradients. While the paper does not 
include a pulsation analysis of any models, the asteroseismic results 
shown in Table~\ref{tab1} suggest that such work could be fruitful.

\subsection*{4.3 More Dimensions}

All of the models we have been discussing so far are 1-D, so they assume 
spherical symmetry. While this is generally a very good assumption for 
white dwarf stars, we {\it know\,} that the effects of rotation and 
magnetic fields can break the spherical symmetry. It has recently become 
computationally feasible to perform star-in-a-box calculations (Turcotte 
et al.~2002), and some of the most important 3-D applications focus on 
core-collapse supernovae (Fryer et al.~2006), which are really just a type 
of white dwarf star hidden in the cocoon of its progenitor. So this is 
another way we might improve the models in the future.

One caveat that I should mention comes from my thesis advisor, Ed Nather.
When asked about the difference between 1-D models and 3-D models, he 
replied: ``3-D models are wrong in {\it three\,} dimensions.''


\section*{5. The Future}

We have discussed two broad approaches to improving our understanding of 
pulsating stars using presently available computational resources. One 
option is to perform a global search by generating {\it millions\,} of 
simple models for comparison with the observations. Or, with similar 
resources, we can calculate a few {\it complete\,} evolutionary tracks 
using relatively sophisticated physical models. My prediction for the 
future should be uncontroversial: as computers get faster, they will 
eventually allow us to combine these two approaches and generate {\it 
millions\,} of {\it complete\,} evolutionary models, opening the door to 
new tests of fundamental physics in pulsating stars.


\acknowledgments{I would like to thank the meeting organizers for inviting 
me to give this review, and for permitting me to write it for a broader 
audience. The National Center for Atmospheric Research is a federally 
funded research and development center sponsored by the U.S. National 
Science Foundation.}

\References{
Althaus, L.~G., et al. 2005, \aap\ 435, 631\\
Beauchamp, A., et al. 1999, \apj\ 516, 887\\
Castanheira, B.~G., et al. 2006, \aap\ 450, 227\\
C{\'o}rsico, A.~H., et al. 2004, \aap\ 427, 923\\
Dantona, F., Mazzitelli, I. 1979, \aap\ 74, 161\\
Elsworth, Y.~P., Thompson, M.~J. 2004, A\&G\ 45, 14\\
Fryer, C.~L., Rockefeller, G., Warren, M.~S. 2006, \apj\ 643, 292\\
Grossman, S.~A., Taam, R.~E. 1996, \mnras\ 283, 1165\\
Iben, I. et al. 1983, \apj\ 264, 605\\
Iben, I.~J., Ritossa, C., Garcia-Berro, E. 1997, \apj\ 489, 772\\
Kanaan, A., et al. 2005, \aap\ 432, 219\\
Kepler, S.~O., et al. 2005, \aap\ 442, 629\\
Metcalfe, T.~S. 2003, \apj\ 587, L43\\
Metcalfe, T.~S. 2005, \mnras\ 363, L86\\
Metcalfe, T.~S., Charbonneau, P.\ 2003, J.~Computat.~Phys.\ 185, 176\\
Metcalfe, T.~S., Nather, R.~E. 2000, Baltic Astronomy, 9, 479\\
Metcalfe, T.~S., Winget, D.~E., Charbonneau, P. 2001, \apj\ 557, 1021\\
Metcalfe, T.~S. et al. 2005, \aap\ 435, 649\\
Mukadam, A.~S., et al. 2004, \apj\ 607, 982\\
Mullally, F., et al. 2005, \apj\ 625, 966\\
Salaris, M., et al. 1997, \apj\ 486, 413\\
Turcotte, S., et al. 2002, ASP Conf.\ 259, 72\\
Unno, W., Osaki, Y., Ando, H., Saio, H., \& Shibahashi, H. 1989, Nonradial\\
\hskip 1.5em oscillations of stars, (Tokyo: University of Tokyo Press)\\
Winget, D.~E., et al.\ 1994, \apj\ 430, 839
}

\end{document}